\documentclass[twocolumn,preprintnumbers,nofootinbib]{revtex4} 
\usepackage{amsmath,amssymb,amsfonts,times,graphicx,color,verbatim,natbib}
\pdfoutput=1

\newcommand{\beq}{\begin{equation}}
\newcommand{\eeq}{\end{equation}}
\def\bea{\begin{eqnarray}}
\def\eea{\end{eqnarray}}

\newcommand{\Sr}{S_{\text{rec}}}
\newcommand{\tr}{\text{tr}}

\begin{document}

\title{Reconstructing quantum states from local data}
\author{Brian Swingle}
\affiliation{Department of Physics, Harvard University, Cambridge MA 02138}
\author{Isaac H. Kim}
\affiliation{Perimeter Institute of Theoretical Physics, Waterloo ON N2L 2Y5, Canada}
\affiliation{Institute for Quantum Computing, University of Waterloo, Waterloo ON N2L 3G1, Canada}

\date{\today}
\begin{abstract}
We consider the problem of reconstructing global quantum states from local data.  Because the reconstruction problem has many solutions in general, we consider the reconstructed state of maximum global entropy consistent with the local data.  We show that unique ground states of local Hamiltonians are exactly reconstructed as the maximal entropy state. More generally, we show that if the state in question is a ground state of a local Hamiltonian with a degenerate space of locally indistinguishable ground states, then the maximal entropy state is close to the ground state projector. We also show that local reconstruction is possible for thermal states of local Hamiltonians. Finally, we discuss a procedure to certify that the reconstructed state is close to the true global state. We call the entropy of our reconstructed maximum entropy state the ``reconstruction entropy", and we discuss its relation to emergent geometry in the context of holographic duality.
\end{abstract}

\maketitle

In this paper we discuss the reconstruction of a global quantum state from local data. As an example, consider a quantum system defined on a one dimensional ring of length $L$ and suppose we are given complete access to all subsystem density matrices of intervals of size $R$. We are furthermore promised that these subsystem states are consistent with a global state.  We would like to know under what conditions the global state can be reconstructed from the local data. Our reconstruction candidate is the global state of maximal global entropy consistent with the local data.  The entropy of the entropy maximizing state is called the reconstruction entropy $\Sr$. Remarkably, when the original global state in question is the unique ground state of a local Hamiltonian with bounded range interactions, perfect reconstruction is possible using our procedure. When the Hamiltonian supports a degenerate ground state manifold of locally indistinguishable states, our procedure reconstructs the projector onto the ground state manifold.

It should be emphasized that the observation that local data determines the ground state for unique ground states of local Hamiltonians is not new \cite{2010NatCo...1E.149C,PhysRevLett.111.020401} (although our reconstruction procedure is different). And of course it is well known that thermal states are maximal entropy states consistent with local energy constraints. Our contributions are refinements of these results, e.g., our demonstration that $\Sr$ provides a certificate guaranteeing closeness of the reconstruction, and the application of the results to open questions in topological order and holographic duality.

We have several motivations for considering this problem. It is known that local data is sufficient to determine many physical quantities of interest in local quantum many-body systems.  In particular, assuming a bounded range Hamiltonian, the energy of the system may be obtained from local data. However, this observation does not lead to an efficient way to find many-body ground states because the problem of determining the existence of a global state consistent with the local data is QMA-complete \cite{2006quant.ph..4166L}. A seemingly different setting in which we would like to understand the relationship between global and local data occurs in gravitational systems.  Holographic duality \cite{1993gr.qc....10026T,susskind,1999IJTP...38.1113M,1998AdTMP...2..253W,1998PhLB..428..105G}, which relates ordinary quantum many-body systems to gravitational systems, shows that these two settings are closely related.

A fundamental question in semiclassical quantum gravity is how to obtain the semiclassical geometrical data from the microscopic degrees of freedom. With general gravitational boundary conditions, this question is still quite mysterious, but using the conjectured holographic duality between gravitational systems in asymptotically AdS spaces and quantum field theories (QFT) associated to the boundaries of those spaces \cite{1999IJTP...38.1113M}, we have a sharp microscopic theory (the QFT) in which to answer such questions for AdS asymptotics.

A clue concerning the diagnosis of the bulk geometry is provided by the observation that surfaces in gravitational theories are often associated with entropies (see e.g., \cite{2012arXiv1212.5183B} for a recent argument). For example, the area in Planck units of the event horizon of a black hole is the entropy of the black hole, although what exactly this entropy represents is unspecified \cite{PhysRevD.7.2333}. We do know that the entropy behaves like thermodynamic entropy, hence a black hole is hot and radiates \cite{Hawking:1974rv}.

Holographic entanglement entropy provides one sharp way to recover geometrical data from entropic quantities \cite{2006PhRvL..96r1602R}. Within Einstein gravity, the entanglement entropy of a field theory region $A$ is given by the bulk area in Planck units of the bulk minimal surface $\tilde{A}$ which is anchored at $\partial A$ at the asymptotic boundary of AdS.  Here we consider a different kind of entropy, the reconstruction entropy $\Sr$, that we use to probe holographic geometry. For highly excited energy eigenstates of a thermalizing holographic Hamiltonian, we show that the reconstruction entropy is associated with the geometry of the corresponding black hole horizon, but for quantum ground states, our results admit no obvious geometrical interpretation.  \cite{2014PhRvD..89h6004B} partially motivated our work by showing that the lengths of bulk curves in AdS$_3$ were related to particular linear combination of entanglement entropies. These combinations of entropies turn out to bound the entropy of the entire system, so \cite{2014PhRvD..89h6004B} conjectured that the lengths of bulk curves were related to some uncertainty in reconstructing the global state from local data.  Our ground state results show that this interpretation of the bulk geometry dual is not always possible if we demand high accuracy reconstruction.

Another important motivation for this work comes from the physics of topological quantum matter (see e.g. \cite{PhysRevLett.48.1559,PhysRevLett.50.1395,doi:10.1142/S0217979290000139,PhysRevB.41.9377}). A crucial question is how to extract the topological properties of the system from feasible measurements. In principle, such topological properties can be obtained, for example, from precisely controlled interferometry experiments involving topological excitations. However, such an approach is prohibitive at present and in any event seems to require considerable knowledge of the detailed physics of the state.  Here we give a very different answer to the question of extracting topological properties by showing that local data suffice to reconstruct the full quantum state provided we consider a ground state or thermal state of a local Hamiltonian.  If the system also possesses translation invariance, then a finite amount of local data suffices to determine the full state and all topological properties. Unfortunately, while we show that there exists a mapping from local data to topological properties, we cannot at present give an efficient way to compute this mapping (which involves constructing the entropy maximizing global state).

Our results are also reminiscent of the Hohenberg-Kohn theorem \cite{PhysRev.136.B864} which forms the basis of density functional theory. One important difference is that we assume a local Hamiltonian throughout, although modifying our arguments to include the Coulomb interaction is straightforward. Note also that our reconstruction results apply to the ground state of any local Hamiltonian and don't require knowledge of the Hamiltonian.

The remainder of the paper is organized as follows. First we define our reconstruction procedure. Then we show that local data permits reconstruction of the ground state projector provided the original global state was the ground state of a local Hamiltonian. This was observed in \cite{2010NatCo...1E.149C} for unique ground states; we show it also for topological states. We further demonstrate that thermal states of local Hamiltonians may also be reconstructed using the maximum entropy method (a subject with a long history, see e.g., \cite{PhysRev.106.620} and more recently \cite{PhysRevLett.111.020401,2012PhRvA..86b2339C,2014arXiv1406.5046C,2013JPhA...46l5301N}) and argue that highly excited energy eigenstates reconstruct to thermal states.  Finally, we discuss reconstruction in the presence of error and show how to certify in certain cases that our reconstruction procedure gives approximately the correct state (without ever knowing said state or the local Hamiltonian).

\textit{Reconstruction entropy.} We begin by defining the problem and the reconstruction entropy $\Sr$. Consider a local many-body system in some (unknown) global state $\rho_{\text{global}}$. We are given complete access to the reduced density matrices $\{\rho_{A_i}\}$ for some set of regions $\{A_i\}$ satisfying $\cup_i A_i = \text{whole system}$. We are promised that this set of density matrices are consistent with at least one global state. Let $\mathcal{C}$ be the set of global states consistent with the local data. This set $\mathcal{C}$ is convex because if $\rho$ and $\rho'$ are both consistent, then so is $p \rho + (1-p) \rho'$ for $p \in [0,1]$. We seek the state $\sigma \in \mathcal{C}$ that maximizes the von Neumann entropy $S(\sigma) = -\text{tr}(\sigma \log(\sigma))$. The maximizer is called $\sigma^\star$ and the maximum entropy is $S(\sigma^\star) = \Sr $.

A general solution to this problem is obtained as follows. Let $\{ O_i \}$ be a complete set of Hermitian operators for the regions $\{A_j\}$ so that their expectation values completely determine the states $\{\rho_{A_j}\}$. For example, if we were considering a single spin, the set $\{I,X, Y, Z\}$ would suffice. More generally, the set of all products of Pauli operators is sufficient for a set of spins. Define
\beq
f(\sigma,\{\lambda_i\}) = S(\sigma) + \sum_i \lambda_i \left(\tr(\rho_{\text{global}} O_i)- \tr(\sigma O_i)\right)
\eeq
where $\rho_{\text{global}}$ is the unknown global state. Since the $O_i$ are restricted to the regions $A_j$, this quantity depends only on local data known to us. Then we consider the variational problem
\beq
\max_{\sigma, \{\lambda_i\}} f(\sigma,\{\lambda_i\}) = f(\sigma^\star,\{\lambda^\star_i\}).
\eeq
The $\lambda_i$ equations of motion impose the consistency with local data, while the variation with respect to $\sigma$ tells us that $\sigma^\star$ has the form
\beq \label{boltzmann}
\sigma^\star = Z^{-1} \exp{\left(- \sum_i \lambda_i^\star O_i\right)}.
\eeq
This is a generalized Boltzmann ensemble familiar from statistical mechanics.

Now we have shown how to construct the entropy maximizing state using only known local data, but it should be emphasized that actually determining the $\lambda_i$ parameters may be a hard computational problem. We must construct the expectation value of every $O_i$ as a function of the $\lambda_i$ and invert the system of equations
\beq
\langle O_i \rangle_\sigma(\lambda) = \langle O_i \rangle_\rho.
\eeq
Such a solution always exists, but it may be hard to find it.

\textit{Reconstruction of local ground states.} We now show that the ground state of a local Hamiltonian can be reconstructed from local data provided the size of the local regions are larger than the range of the local Hamiltonian. The result is easiest for states which are unique ground states of local Hamiltonians; later we generalize the argument to include degenerate locally indistinguishable ground states. The results are illustrated with an explicit free fermion example in Appendix 1.

Suppose the unknown global state $\rho$ is the unique ground state of some local Hamiltonian $H$ with range $R_0$. $H$ is a sum of geometrically local terms,
\beq
H = \sum_x H_x,
\eeq
such that each $H_x$ acts only on degrees of freedom within $R_0$ of $x$. We will need not to know anything about $H$ in what follows except that it exists and has range $R_0$.

Now let $\sigma^\star$ be the global state of maximal entropy consistent with local data on regions of size $R$. Following the discussion in Section 2, we can construct this state as an exponential of sums of local operators defined on the same regions of size $R$. Determining $\sigma$ is then a matter of choosing the $\lambda$ parameters such that the global state is consistent with local data. This may be a hard computational problem in general, but in principle the state $\sigma$ can be determined.

One locally consistent solution is
\beq
\sigma^\star = \lim_{\beta \rightarrow \infty} e^{-\beta H},
\eeq
in which case the reconstruction entropy is just zero. As we now show, this is the unique answer.

Let the energy of the ground state $\rho$ be $E_0$ and consider the positive operator $H - E_0 \geq 0$. We must have
\beq
\tr(\sigma^\star (H-E_0)) \geq 0
\eeq
by positivity. However, using the locality of $H$ we have
\beq
\tr(H_x \sigma^\star) = \tr(H_x \rho)
\eeq
provided $R \geq R_0$ since $\rho$ and $\sigma^\star$ agree on sub-systems of size $R$ and smaller. Hence we may write
\beq
\tr(\sigma^\star (H-E_0)) = \tr(\rho(H-E_0)) = 0,
\eeq
where the last equality follows because $\rho$ is the ground state. We finally conclude that
\beq
\tr(\sigma^\star (H-E_0)) = 0,
\eeq
and since $(H-E_0)$ is positive $\sigma^\star$ must lie in its null space. But by assumption $\rho$ was the unique ground state of $H$, hence $\sigma^\star = \rho$ as claimed.

The above discussion can be applied to topological states as well. The new feature which appears is the possibility of degenerate ground states which are locally indistinguishable. For example, consider a $Z_2$ gauge theory (e.g. a spin liquid \cite{PhysRevLett.66.1773}) in the extreme deconfined limit, e.g. with no string tension. On a torus, there are four states which are locally identical but differ in their values for certain non-local string operators. The maximum entropy state consistent with local data is then the equal weight mixture of the four ground states. More generally, this result holds for any model with ground states which are exactly locally indistinguishable. Since the statistics and braiding of excitations can be extracted from a complete set of ground states on a torus \cite{PhysRevB.85.235151}, we conclude that the same information can be extracted from the local density matrices as well. As we discuss in Appendix 2 and 3, the above argument remains valid even if the states are only approximately locally identical.

\textit{Reconstruction of thermal states.} We now extend the results of the previous section to encompass thermal states of local Hamiltonians. We show that any thermal state of a local Hamiltonian can be reconstructed from local data using the maximum entropy state discussed above. Note that this idea is very old, e.g., \cite{PhysRev.106.620} used it extensively, but extra care is required when dealing with topological systems or allowing errors. The ground state results of the previous section are obtained as a limit of the results in this section.

Consider again a local Hamiltonian $H$ and suppose the system is in a thermal state at temperature $T$,
\beq
\rho = \frac{1}{Z} e^{- H/T}.
\eeq
The thermal state is characterized as maximizing $S(\rho)$ subject to the constraint that $\langle H \rangle_\rho = E$ where $E$ is some fixed energy. The temperature is a Lagrange multiplier, e.g. we extremize $S(\rho) - \beta (\langle H \rangle - E)$ with respect to both $\rho$ and $\beta$ and find $\beta = 1/T$.  Hence the thermal state is the maximum entropy state consistent with just one constraint on the total energy.

Let $\sigma^\star$ be the maximum entropy state consistent with $\rho$ on regions of size $R$. If $R > R_0$, the range of $H$, then by the assumption of local consistency $\sigma^\star$ correctly computes the expectation value of all the local terms in $H$ and hence correctly computes the expectation value of $H$ itself. Thus $\sigma^\star$ is among the states consistent with the total energy constraint, hence $S(\rho) \geq S(\sigma^\star)$ since $\rho$ is the entropy maximizing state consistent with the total energy constraint.

To complete the argument, we introduce the relative entropy $S(\rho|\sigma)$ (see \cite{RevModPhys.50.221} for a review) defined as
\beq
S(\rho|\sigma) = \tr(\rho \log(\rho) - \rho \log(\sigma)).
\eeq
The relative entropy is not symmetric in its argument, but it does provide a kind of quasi-distance between states. Furthermore, the relative entropy obeys $S(\rho |\sigma) \geq 0$ and is only zero if $\rho = \sigma$. It can also be infinite if the support of $\sigma$ is smaller than the support of $\rho$.

From the definition we have
\beq
S(\rho | \sigma^\star) = \tr(\rho \log(\rho) - \rho \log(\sigma^\star)) = - S(\rho) - \tr(\rho \log(\sigma^\star)).
\eeq
By construction $\log(\sigma^\star)$ is a sum of local operators with each local operator supported on a region where $\sigma^\star$ is consistent with $\rho$. Hence we may write
\beq
\tr(\rho \log(\sigma^\star)) = \tr(\sigma^\star \log(\sigma^\star)) = - S(\sigma^\star).
\eeq
The relative entropy is then
\beq
S(\rho | \sigma^\star) = S(\sigma^\star) - S(\rho),
\eeq
so we conclude from positivity that $S(\sigma^\star) \geq S(\rho)$.

Combining the two inequalities $S(\rho)\geq S(\sigma^\star)$ and $S(\sigma^\star) \geq S(\rho)$ we must have $S(\sigma^\star) = S(\rho)$. Then the relative entropy is zero, so $\rho = \sigma^\star $ as claimed. Note that we do not have to know $H$, only that $H$ exists, to prove that $\rho = \sigma^\star$.

\textit{Reconstruction with error.} We now modify our results to allow reconstructed states that only reproduce local data up to some error. It is important for our purposes that we don't modify the local data itself since this could lead to inconsistent local data. One motivation for introducing error is that perfect local consistency is never really achieved since there are always experimental uncertainties. Reassuringly, we show that the constructions above tolerate small errors in the local consistency conditions. For example, demanding local consistency up to an error of order inverse polynomial in the system size still gives a reconstructed state which is close to the target state.

Suppose $\rho$ is the unique ground state of a local range $R_0$ Hamiltonian $H$ consisting of $L^D$ terms in $D$ dimensions. Let $\sigma^{\star,\epsilon}$ be the maximum entropy state of the form Eq. (\ref{boltzmann}) with the property that
\beq
\|\rho_R - \sigma^{\star,\epsilon}_{R} \|_1 \leq \epsilon.
\eeq
We again compute the quantity $\tr(\sigma^{\star,\epsilon}(H-E_0)) \geq 0$ and find
\beq
\tr(\sigma^{\star,\epsilon}(H-E_0)) \leq \epsilon L^D \max_x(\|H_x\|).
\eeq
If $\Delta$ is the gap of $H$ then we also have
\beq
\tr(\sigma^{\star,\epsilon}\Delta (1-\rho)) \leq \tr(\sigma^{\star,\epsilon}(H-E_0)),
\eeq
where $(1-\rho)$ projects onto the orthogonal complement of the ground state. If $\Delta \sim \frac{1}{\text{poly}(L)}$ then $\epsilon \sim \frac{1}{\text{poly}(L)}$ is sufficient to obtain high overlap with the true ground state $\rho$.

The thermal argument is also quite similar to the case of perfect reconstruction. However, there is one important subtlety: the need to control the difference between $\tr(\rho \log(\sigma^\star))$ and $ \tr(\sigma^\star \log(\sigma^\star))$ places a restriction on the lowest temperatures we can consider for a given error $\epsilon$. For any temperature independent of system size (or slowly decreasing, e.g. $T\sim 1/\log(L)$), $\frac{1}{\text{poly}(L)}$ error is sufficient to reconstruct the thermal state to high accuracy.

To make contact with the geometry of black holes we turn to the case of highly excited energy eigenstates of a thermalizing Hamiltonian $H$. Assuming a strong form of eigenstate thermalization \cite{PhysRevA.43.2046,1994PhRvE..50..888S}, it follows that for given energy eigenstate $|E\rangle$ and a small region $A$ we have
\beq
\tr_{\bar{A}}(|E\rangle \langle E|) \approx \tr_{\bar{A}}\left(e^{-H/T(E)}/Z\right).
\eeq
In the language of the trace norm $\| ...\|_1$, we have (with high probability or for almost all states)
\beq
\left\| \tr_{\bar{A}}(|E\rangle \langle E|) - \tr_{\bar{A}}\left(e^{-H/T(E)}/Z\right) \right\|_1 \leq \epsilon.
\eeq
In this equation $\epsilon$ can be exponentially small in the total system size \cite{2006NatPh...2..754P}.

Thus allowing a very small amount of error in the local data extracted from a highly excited state immediately precludes the possibility of perfect reconstruction. Furthermore, we have shown that thermal states are stable points in that they can be perfectly reconstructed, so the maximal entropy state obtained from a highly excited state will be very close in trace norm to the corresponding thermal state at temperature $T(E)$ determined by $E$. Suppose the system in question is also a holographic QFT so that the thermal state is dual to a black hole (BH) geometry.  Then the reconstruction entropy obeys
\beq
\Sr = S_{\text{thermal}} = S_{\text{BH}} = \frac{A_{\text{BH}}}{4 G_N},
\eeq
where $G_N$ is Newton's constant.  Hence the reconstruction entropy computes the area $A_{\text{BH}}$ of the dual black hole horizon. This result provides an interesting geometrical interpretation of the reconstruction entropy of a highly excited pure state.

\textit{Certifying the reconstruction.} So far we have outlined a general reconstruction procedure that is applicable to any quantum state. We discussed several examples and showed that the procedure works well in a variety of physically relevant scenarios.

Now we ask whether one can certify that the reconstructed state is close to the original state without knowing the original state or the Hamiltonian. We show that this is possible provided that the reconstruction entropy is close to zero.

To make things concrete, suppose we are given a quantum state $\rho$. Recall that we defined $\sigma^{\star,\epsilon}$ as the maximum entropy state that is approximately consistent with $\rho$ over local subsystems with a precision $\epsilon.$ The entropy of $\sigma^{\star,\epsilon}$ is $S_{\text{rec}}^{\epsilon} = -\tr(\sigma^{\star,\epsilon}\log(\sigma^{\star,\epsilon}))$.

We prove a universal upper bound on the distance between $\rho$ and $\sigma^{\star,\epsilon}$:
\begin{equation}
\frac{1}{8}\|\rho - \sigma^{\star, \epsilon}\|_1^2 \leq S_{\text{rec}}^{\epsilon} - S(\rho),\label{eq:certificate}
\end{equation}
which follows from a general inequality between two quantum states, $\frac{1}{8}\|\rho - \sigma \|_1^2 \leq S(\frac{\rho+\sigma}{2}) - \frac{S(\rho)+S(\sigma)}{2}$ \cite{2013arXiv1310.0746K}. Since $\sigma^{\star,\epsilon}$ is the maximum entropy state, $S(\frac{\rho+\sigma^{\star,\epsilon}}{2}) \leq S_{\text{rec}}^{\epsilon}$ and $S(\rho) \leq S_{\text{rec}}^{\epsilon}$. Plugging in these two inequalities, Eq. (\ref{eq:certificate}) is derived. This inequality assigns an operational meaning to the reconstruction entropy, since its smallness certifies the faithfulness of the reconstruction procedure.

\textit{Outlook.} Motivated by questions in topological quantum matter and holographic duality, we studied the problem of reconstructing global states from local data. We gave an explicit procedure to reconstruct the state by considering the maximum entropy state which is consistent with the local data. This procedure may be computationally difficult to carry out in general, but in some cases finding the maximal entropy state is not too difficult. In any event, we leave for future work precise statements about the complexity of the reconstruction process.

One context where reconstruction is easy to carry out occurs when the target state is a quantum Markov chain. Quantum Markov chains are states that saturate strong sub-additivity of entropy and hence have a very special conditional structure. Such states have been characterized in \cite{2004CMaPh.246..359H}, and Petz has shown that there is a quantum channel which permits one to reconstruct a Markov chain piece by piece \cite{petz1986}. Some gapped ground states of local Hamiltonians are Markov chains or nearly Markov chains, so we suspect that this technology will be useful for further work in the reconstruction problem. Related ideas have already been pursued in \cite{2011PhRvL.106h0403P}.

One idea for making further contact with holographic geometry considers the reconstruction problem for subsystems. Suppose we reconstruct not the global ground state of a holographic QFT but only the state of a sub-region in the ground state. Then the reconstruction entropy will be related to the entanglement entropy which has a geometrical interpretation. For example, in the ground state of a CFT \cite{2011JHEP...05..036C} has shown that the density matrix of a ball in any dimension is the exponential of a local operator. The result for thermal states of local Hamiltonians then implies that exact reconstruction is possible for this region type. Hence the reconstruction entropy is the entanglement entropy which is geometrical.

Finally, turning to our topological phases motivation, we have shown that the ground state projector in a topological phase can be reconstructed from local data. In principle this gives a map from local data to topological data. A better understanding of the properties of this map, e.g. if it is efficiently computable, is an interesting target for future work.

We thank N. Lashkari and P. Hayden for useful conversations. BGS is supported by a Simons Fellowship through Harvard University and thanks the Simons Center for the Theory of Computation at Berkeley, the Perimeter Institute, and Stanford University for hospitality during this work. IK's research
at Perimeter Institute is supported by the Government of Canada through Industry
Canada and by the Province of Ontario through the Ministry of Economic Development
and Innovation.

\bibliography{reconstruction_v2}

\appendix

\subsection{Explicit example with free fermions}

Here we give an explicit example in a gapless free fermion system. To simplify the discussion, we will assume that all states obey Wick's theorem. Consider fermions defined on a ring with $L$ sites and with Hilbert space generated by the fermion operator $c_r$ with $r=1,...,L$. Suppose the Hamiltonian has the form
\beq
H = - w \sum_{r} c_{r+1}^\dagger c_r + h.c.
\eeq
so that we have two Fermi points. The dispersion is $E_k = - 2 w \cos{(k)}$, and the fermion equal time two-point function is
\beq
G_{x,y} = \langle c_x^\dagger c_y \rangle = \frac{1}{L} \sum_k e^{ik(x-y)} \theta(-E_k).
\eeq

Now suppose we only have access to $G$ on local patches of size $R$. In other words, we know $G_{x,y}$ for all $x$ and $y$ satisfying $|x-y|\leq R$. The entropy maximized state consistent with this data then has the form
\beq
\sigma = Z^{-1} \exp{\left( - \sum_{r=1}^L \sum_{\delta = 0}^{R} \lambda_{r,\delta} (c_{r+\delta}^\dagger c_{r} + h.c.) \right)}.
\eeq
Translation invariance of the local data forces $\lambda_{r,\delta}$ to be independent of $r$. Hence we have just $R+1$ couplings to determine. The operator in the exponential has a spectrum given by
\beq
\tilde{E}_k = \sum_{\delta=0}^\ell 2 \lambda_{\delta} \cos{(\delta k)},
\eeq
and hence the fermion correlator obtained from it has the form
\beq
\tilde{G}_{x,y} = \frac{1}{L} \sum_k \frac{e^{i k(x-y)}}{e^{\tilde{E}_k}+1}.
\eeq

Start with $R=0$. Then we know only $G_{0,0}$ which is the density of particles $n = 1/2$. The density of particles in the reconstructed state is
\beq
\tilde{n} = \frac{1}{L} \sum_k \frac{1}{e^{2 \lambda_0}+1} = \frac{1}{e^{2 \lambda_0}+1},
\eeq
so $\lambda_0$ is determined by $n = \tilde{n}$ to be
\beq
\lambda_0 = 0.
\eeq
The entropy of the reconstructed state is $\Sr = L \log(2)$.

Now consider $R=1$. We know $G_{0,0}$ and $G_{0,1}$ which are
\beq
G_{0,0} = \frac{1}{2}
\eeq
and
\beq
G_{0,1} = \int_{-\pi/2}^{\pi/2} \frac{dk}{2\pi} e^{ik} = \frac{e^{i\pi/2} -e^{-i\pi/2}}{2\pi i} = \frac{1}{\pi}.
\eeq
The reconstructed Greens function is
\beq
\tilde{G}_{x,y} = \int \frac{dk}{2\pi} \frac{e^{i k(x-y)}}{e^{\tilde{E}_k}+1}
\eeq
with $\tilde{E}_k = 2 \lambda_0 + 2 \lambda_1 \cos{(k)}$.  We have
\beq
\tilde{G}_{0,0} = \int \frac{dk}{2\pi} \frac{1}{e^{\tilde{E}_k}+1}
\eeq
and
\beq
\tilde{G}_{0,1} = \int \frac{dk}{2\pi} \int \frac{dk}{2\pi} \frac{e^{i k}}{e^{\tilde{E}_k}+1}.
\eeq
By symmetry we must have $\lambda_0 = 0$ to give the correct density. It then follows from a simple analysis that the only locally consistent solution is $\lambda_1 = - \infty$ (which is the exact ground state).

The reconstruction entropy is a non-increasing function of $R$ because we add more constraints as $R$ increases. Indeed, any state which is consistent at size $R'$ is also consistent for all $R < R'$, so the entropy maximizing state at $R'$ is a candidate for the entropy maximizing state at $R < R'$, and hence $\Sr(R) \geq \Sr(R')$ if $R' \geq R$. Since $\Sr(R=1) = 0$, it follows that $\Sr = 0$ for all $R \geq 1$. Thus exact reconstruction is indeed possible in this free fermion example and is even easy to perform.

\subsection{Robust reconstruction of the locally indistinguishable ground state subspace}
We discuss a subtlety which was omitted in the discussion on topological states. Namely, realistic topological ground states are unlikely to be exactly locally indistinguishable from each other. Typically there is a small but nonzero discrepancy that decays exponentially with the system size. We show that the reconstruction still works in such cases.

Without loss of generality, denote $\{|\psi_i \rangle \}_{i=1,\cdots,N}$ as a set of locally indistinguishable ground states. The local indistinguishability condition can be stated as follows:
\begin{align}
|\langle \psi_i | O |\psi_j \rangle| &\leq \delta \,\,\, (i \neq j) \\
|\langle \psi_i |O | \psi_i \rangle - \langle \psi_j |O | \psi_j \rangle | &\leq 2\delta,
\end{align}
where O is an arbitrary normalized local operator. Typically $\delta$ decays exponentially with the linear size of the system, which is denoted as $L$. Let us assume that the spectral gap, denoted as $\Delta$, remains finite in the thermodynamic limit.

Under these assumptions, the parent Hamiltonian of the ground states can be decomposed into two parts:
\begin{equation}
H=H_{\text{low}}+H_{\text{high}}.
\end{equation}
Here, $H_{\text{low}}$ is the low energy part of the Hamiltonian, which consists of eigenstates having energy lower than $\Delta$. The other term is the remaining part of the Hamiltonian. From the spectral condition, the weight of $\sigma^{\star,\epsilon}$ in the high energy sector can be bounded:
\begin{equation}
 \text{Tr}(P_{\text{high}} \sigma^{\star,\epsilon})\leq \frac{\text{Tr}(H \sigma^{\star,\epsilon})}{\Delta}.
\end{equation}
By the local consistency condition, the energy of $\sigma^{\star,\epsilon}$ can be bounded from above by $\epsilon L^{D}$ in a $D$-dimensional system. Combining these two, the weight on the high energy sector is bounded by $\frac{\epsilon L^D }{\Delta}$. Using the concavity of entropy,
\begin{equation}
S(\sigma^{\star,\epsilon}) \leq S(p_{\text{low}}\sigma^{\star,\epsilon}_{\text{low}} + p_{\text{high}}\sigma^{\star,\epsilon}_{\text{high}}),
\end{equation}
where $p_{\text{low}}=\text{Tr}(P_{\text{low}} \sigma^{\star,\epsilon})$, $\sigma^{\star,\epsilon}_{\text{low}} = P_{\text{low}} \sigma^{\star,\epsilon}P_{\text{low}} / p_{\text{low}}$, and the other terms are defined similarly. Using the continuity of entropy \cite{fannes},
\begin{equation}
|S(p_{\text{low}}\sigma^{\star,\epsilon}_{\text{low}} + p_{\text{high}}\sigma^{\star,\epsilon}_{\text{high}}) - S(\sigma^{\star,\epsilon}_{\text{low}})| \leq 2p_{\text{high}} L^D\log \frac{d}{2p_{\text{high}}},
\end{equation}
where $d$ is the dimension of the particles. Since $S(\sigma^{\star,\epsilon}_{\text{low}})$ is at most $\log N$, we conclude that, for topological states,
\begin{equation}
\frac{1}{8}\|\rho - \sigma^{\star,\epsilon} \|_1^2 \leq 2p_{\text{high}}L^{D} \log \frac{d}{2p_{\text{high}}},
\end{equation}
where $p_{\text{high}}\leq \frac{\epsilon L^D}{\Delta}$. For an $\epsilon \geq \delta$ that decays sufficiently fast in $L$, the bound converges to $0$ in the thermodynamic limit.

\subsection{Thermal argument for approximate locally indistinguishable states}

We can also reproduce the results of Appendix 2 using our thermal results and an assumption about the low temperature free energy. Consider a topological phase with a Hamiltonian $H$ which has a gap $\Delta$ separating approximately degenerate ground states $\{|\psi_i \rangle \}_{i=1,\cdots,N}$ from the rest of the spectrum. The ground states satisfy the local indistinguishability conditions
\begin{align}
|\langle \psi_i | O |\psi_j \rangle| &\leq \delta \,\,\, (i \neq j) \\
|\langle \psi_i |O | \psi_i \rangle - \langle \psi_j |O | \psi_j \rangle | &\leq 2\delta,
\end{align}
with $O$ any normalized local operator and $\delta$ exponentially small in system size $L$.

Now consider the thermal state
\beq
\rho(T) = Z^{-1}(T) e^{-H/T},
\eeq
and suppose the low temperature free energy $F = - T \log(Z)$ obeys $F = - T \log(N) + F_{\text{excited}}$ with $F_{\text{excited}} \sim L^D e^{-\Delta/T}$ up to corrections which are exponentially small in system size.  Let the projector onto the ground state space be $P = \sum_i |\psi_i \rangle \langle \psi_i|$. The difference in trace norm between $P/N$ (the equal weight mixture of ground states) and $\rho(T)$ is
\beq
\left\| \frac{P}{N} -\rho(T) \right\|_1 = \frac{1}{N} - \frac{1}{Z(T)} + \frac{Z(T)-N}{Z(T)}
\eeq
up to terms exponentially small in system size.

The assumed form of $F$ then guarantees that by taking $T \sim \frac{\Delta}{\log(L)}$ we may make $\left\| \frac{P}{N}-\rho(T)\right\|_1 \sim \frac{1}{\text{poly}(L)}$. We have thus shown that there is a thermal state which is close to the ground state projector and approximately reproduces local properties any ground state. Since we lack the exponential accuracy necessary to distinguish the ground states, it follows that the maximal entropy state approximately consistent with local data from any pure state in the ground state manifold is close to the projector onto the ground state manifold.

\end{document}